\documentclass{article}
\usepackage[margin=1.2in]{geometry}
\usepackage[utf8]{inputenc}

\usepackage{graphicx}
\usepackage{hyperref}
\usepackage{enumitem}
\usepackage[affil-it]{authblk}

\title{Assessing Disease Exposure Risk with Location Data: A Proposal for Cryptographic Preservation of Privacy}

\author[1]{Alex Berke}
\author[1,2]{Michiel Bakker}
\author[1]{Praneeth Vepakomma}
\author[1]{Kent Larson}
\author[1,2]{Alex `Sandy' Pentland}
\affil[1]{MIT Media Lab, Cambridge, MA, USA}
\affil[2]{MIT Connection Science, Cambridge, MA, USA}
\affil[ ]{\quad}
\affil[ ]{\normalfont{\texttt {\{aberke,bakker,vepakom,kll,pentland\}@mit.edu}}}

\date{March 2020}

\begin{document}

\maketitle

\begin{abstract}
    Governments and researchers around the world are implementing digital contact tracing solutions to stem the spread of infectious disease, namely COVID-19. Many of these solutions threaten individual rights and privacy. Our goal is to break past the false dichotomy of effective versus privacy-preserving contact tracing.  We offer an alternative approach to assess and communicate users' risk of exposure to an infectious disease while preserving individual privacy. Our proposal uses recent GPS location histories, which are transformed and encrypted, and a private set intersection protocol to interface with a semi-trusted authority. 

    There have been other recent proposals for privacy-preserving contact tracing, based on Bluetooth and decentralization, that could further eliminate the need for trust in authority.
    However, solutions with Bluetooth are currently limited to certain devices and contexts while decentralization adds complexity. The goal of this work is two-fold: we aim to propose a location-based system that is more privacy-preserving than what is currently being adopted by governments around the world, and that is also practical to implement with the immediacy needed to stem a viral outbreak.

\end{abstract}

\section{Introduction}
This is a proposed design for how a system can identify and notify users who may have come in contact with diagnosed disease carriers about their risk while respecting individual' privacy. It is common practice for healthcare workers to interview individuals diagnosed with a communicable disease about their recent movements in order to identify and alert others who they may have come in contact with.  This process is referred to as {\em contact tracing} and is crucial as health entities, communities and governments attempt to contain viral outbreaks. Manually performing contact tracing is highly resource-intensive, intrusive and time-consuming.  Yet the ubiquitous use of personal digital devices provides access to detailed location histories that are collected automatically, enabling a system that could conduct this process at scale. The objective of this paper is to describe such a system in the face of the COVID-19 pandemic that better preserves the privacy of its users than the systems we see governments adopting.

\subsection{Contact tracing \& privacy}
There currently are digital approaches to contact tracing that use location histories\footnote{There are also digital contact tracings solutions that use Bluetooth technology to detect points of contact between individuals without using location histories directly~\cite{lewis2020coepi,singapore2020trace,covidwatch2020}.  We will discuss these in Section \ref{sec:alternative}.} but many operate on a skewed trade-off between privacy and effectiveness~\cite{raskar2020apps}. Some rely on general public broadcasting of information that introduces uncertainty in the information disseminated.\footnote{Information about where diagnosed COVID-19 carriers have visited has been publicly broadcasted in  different ways. Singapore updates a map and lists details about the identities about each COVID-19 case~\cite{singaporeCases}. South Korea sends text messages to citizens containing personal information about diagnosed carriers \cite{zastrow2020south}. In the US, location information about diagnosed patients has been published through media outlets and university and government websites~\cite{nytimesCoronaVirusCount,johnhopkinsCOVID19cases,nebraskaCOVID19}.} Other alternatives resort to the usage of technologies that risk violating individual rights against stigmatization and surveillance\footnote{Israel employs cellphone tracking for coronavirus patients~\cite{haaretzIsraelApprovesTracking}.  China requires use of an COVID-19 mobile application  to effectively surveil citizens and dictate quarantines~\cite{nytimesChinaColorCode}.} (see~\cite{raskar2020apps} for a discussion).

Even when systems do attempt to use location data while respecting individual privacy, their methods are often insufficient: simply anonymizing user location histories by replacing users' identifiers with random new ones fails to achieve privacy in a meaningful way. Secondary data that is not anonymized can be used to re-identify users by matching data points across the datasets.  For example, a study using credit cards records has shown that only four data points from a user's location history are enough to uniquely re-identify 90\% of individuals \cite{de2015unique}.

This work breaks past the dichotomy of privacy versus effectiveness established by previous digital contact tracing approaches.  We propose a technology-based solution for contact tracing and disseminating information on the state of infections while protecting the privacy rights of both diagnosed infected carriers and other citizens.

\subsection{Trust \& privacy principles}
The privacy and trust principles central to the design of this work are summarized below:
\begin{itemize}
    \item \textbf{Keep location data private.} Locations visited are kept private for all users including those who are diagnosed disease carriers.
    \item \textbf{Avoid surveillance.} The system can detect points of contact between users without precise location histories being exposed.
    \item \textbf{Only allow one-way private data publication.} Only diagnosed carriers ever publish data, but this data remains encrypted and private, and their identities remain protected. Other users can check if they came in contact with carriers without sharing their location histories.
\end{itemize}

Any privacy-preserving contact tracing framework should be considered a ``best effort'' and avoid promising to be perfectly private.  Our primary contribution to the space of existing frameworks and digital tools is the degree to which our cryptographic approach can preserve user privacy while providing highly useful and accurate information through individuals GPS location histories.     

We present our proposal in response to current events where we see an increase in surveillance and a forfeit of privacy due to contact tracing efforts designed to contain the spread of COVID-19.  We intend to show that effective contact tracing is possible without further forfeiting privacy.

\subsection{A useful first step}
A simple first version of a system that provides exposure risk information is one that collects, anonymizes, and aggregates the recent GPS location histories of diagnosed carriers. This information allows the creation of a spatiotemporal \emph{heatmap} representing large geographic regions where diagnosed carriers spent time and when. 

Individuals' data and areas visited must be aggregated and obfuscated in a way that minimizes what can be learned about individual people or places visited in the dataset. This is done in order to protect people and businesses from potential stigmatization or any other threat.  This aggregated view can provide helpful information about infection risk across different areas, types of places, and time periods for both health authorities and the general public. This aggregated data can be further analyzed to better understand the flow and trends of disease transmission.

\subsection{Contributions}
In the current paper we build upon this first step and discuss the potential use of a {\em private set intersection} protocol
to provide more precise risk assessments to individual users based on points of contact with individuals who were later diagnosed as disease carriers.
Our approach partitions the space of fine-grained GPS location and time data into discrete spatiotemporal points that represent location histories.
This combination of a partition  scheme and private set intersection protocol allows the system to detect when a user came in contact (e.g. was in close proximity) with diagnosed carriers to assess and inform them of their risk, while preserving the privacy of individuals.

\section{A Short Survey of Alternative Solutions}\label{sec:alternative}
Clever approaches to privacy-preserving systems for contact tracing that rely on Bluetooth rather than GPS have been proposed and implemented as smartphone applications, such as TraceTogether \cite{singapore2020trace}, which was created by the government of Singapore, and the  open source projects COVID Watch \cite{covidwatch2020} and CoEpi \cite{lewis2020coepi}.  Instead of recording user location histories, these systems record when users come in proximity of one another.  A user's smartphone application generates, and periodically updates, a random identifier called a ``contact event number'' (CEN).  The user's device broadcasts this CEN using Bluetooth Low Energy and receives and locally records CENs broadcast by other users' devices.  Users who later fall ill then share their CENs, which can be matched against the CENs recorded locally by other users to identify points of contact between diagnosed users and the other users.
The different implementations vary in the privacy and security they provide their users. For example, the Singapore government requires users of its TraceTogether app to share their phone numbers and identities with a central authority that maintains a database of recorded CENs. This authority contacts users whose recorded CENs match against diagnosed users' CENs to alert them of their exposure. This system provides users with no privacy from the  central authority but does protect users from having their identities exposed to other users. In contrast, the COVID-Watch and CoEpi projects place more trust in users rather than in a central authority.\footnote{The CoEpi project allows users to self-diagnose and share their CENs without a health official verifying their diagnosis and data.  This can result in false reporting and invites spam attacks from malicious users, which can result in other users seeing false positives which, in turn, could cause panic or simply distrust of the system. This issue has been acknowledged by CoEpi's technical leadership and they have invited solutions from researchers \cite{niyogi2020memo}.} With their designs, diagnosed users' CENs  are stored  in a database that is either made public for other users' apps to download or made available for them to poll \cite{niyogi2020memo,niyogi2020github}. These models are not perfectly privacy-preserving either, as they are susceptible to privacy attacks from other users, as described by Cho et al. \cite{cho2020contact}.

These systems use Bluetooth to directly detect whether users come in proximity of one another, which GPS cannot provide. With Bluetooth, proximity can be approximated by signal strength that is reduced by obstructions like walls; therefore, it more accurately reflects functional proximity in high-risk environments for close contact, such as within buildings and vehicles, or in underground transit.  These Bluetooth features can benefit our proposed system as well, as the installation of Bluetooth beacons could mark locations with higher levels of accuracy than what would otherwise be marked only with GPS coordinates.

We note that applications that detect users' exposure risk by relying on users to come into Bluetooth proximity may not be sufficient, given that in addition to human-to-human transmission, coronaviruses can transmit through commonly touched surfaces or environments~\cite{kampf2020persistence}. In contrast, our proposed system can be extended to handle this kind of exposure risk by checking adjacent time periods to accommodate when two users may have occupied the same space one after another.

Another issue with purely Bluetooth-based systems is that they require people to use a special application that could suffer from slower or limited adoption, limiting the ability to capture a large enough base of users before they are diagnosed, and therefore limiting their potential to be useful. With GPS based approaches we might leverage the fact that the applications on users' phones are already collecting their GPS location histories, and develop a technology as a plugin to interface with these apps, to more quickly roll out our system as a life-saving contact tracing technology.

\section{A GPS-based privacy-preserving scheme}
In what follows, we describe what type of information our proposed system provides before showing a high-level system overview.  We then explain the trust and privacy model it is designed for, and finally provide a more technical description with details on how the system could be built in practice. 

\subsection{Probabilistic risk assessment
}
The proposed system provides a probabilistic measure of disease exposure risk for a user, based on the time they have spent in spaces shared with users who were later diagnosed as disease carriers.  More time spent in such shared spaces indicates higher levels of risk,  but this risk is also dependent on where those spaces are.

Any technological system should be wary of claiming to precisely determine exposure risk, due to the limitations of the technologies used for detection, and the ambiguity over what types of interactions between people, shared spaces, or common surfaces, most elevate risk.   Our proposed system uses GPS points collected from users' devices and can be extended to use Bluetooth to indicate locations visited as well. It is worth noting that GPS has limited accuracy,  especially in dense urban environments or multi-story buildings. But even detecting whether a user spent sustained time in a crowd or multi-story building with a diagnosed carrier may be useful, due to the heightened likelihood of sharing not only space but surfaces such as door handles or elevator buttons, which help a virus spread \cite{kampf2020persistence}. For this reason, we propose a probabilistic risk assessment based on the amount of time and number of places that a user shared with a disease carrier,  which we call ``points of contact''.

\subsection{System overview}
Below we provide a high-level overview of our system and walk through an illustrated example.

\begin{figure}
    \centering
    \includegraphics[width=0.8\textwidth]{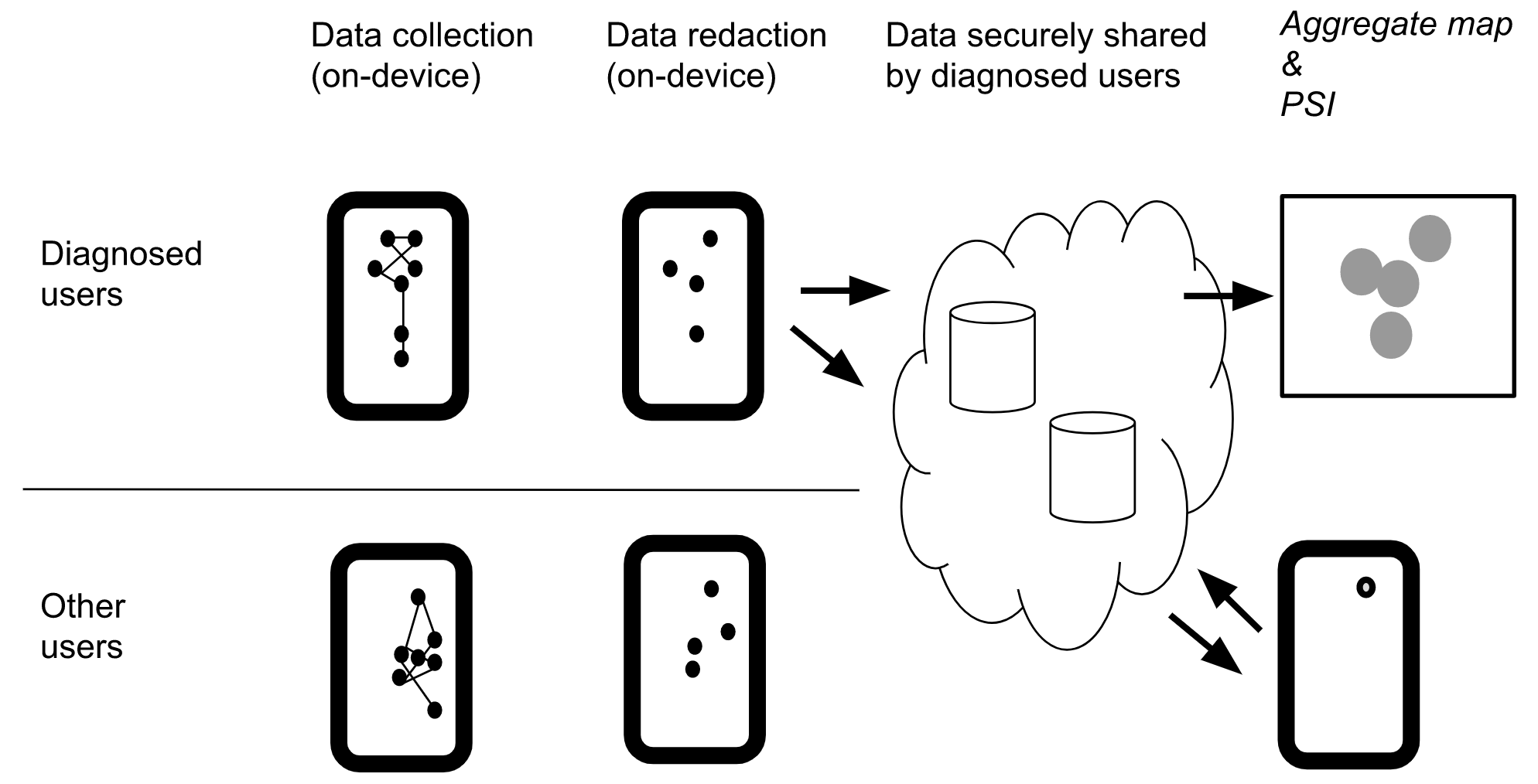}
    \caption{High-level schematic showing the major steps in the system's process for privacy-preserving contact tracing:
    (1) data collection, (2) redaction and transformation of data,
    (3) secure data exchange, and 
    (4) individualized risk assessment and notification, as well as the distribution of aggregated data to create a `heatmap' to inform the public of more general risk.
    A more detailed description of the process includes usage of discrete spatiotemporal `point intervals' and `a private set intersection protocol and is described later on in this paper.}
    \label{fig:high_level_schematic}
\end{figure}

Our proposed system follows four (4) steps, namely 
(1) data collection, 
(2) redaction and transformation of data,
(3) secure data exchange, and 
(4) risk assessment and notification.  We expand upon these steps below (see Figure~\ref{fig:high_level_schematic}).

\begin{description}

\item	[Step~1:] {\bf Data collection}.
An application (app), installed on the mobile device of the user,
collects timestamped GPS points throughout the user's day, every $t$ minutes. 
The sequence of points represents their location history.

\item	[Step~2:] {\bf Redaction and transformation of data}.
All collected location histories are redacted, transformed, and encrypted before leaving the device, to protect user privacy. In the case that a user is diagnosed, their points are transformed and shared 
with the server in a way that maintains their privacy. 
Collected points are deleted from both devices and servers $d$
days after they were collected, where $d$ is the period of possible 
disease transmission informed by medical experts.

\item	[Step~3:] {\bf Secure data exchange}.
In this phase, using the agreed secure data exchange mechanism,
the mobile app (acting as the client) establishes a secure channel
with the designated server.
Within this secure channel,
the mobile application requests the server
for the `point interval' data of known infected carriers for a chosen duration of time (e.g. the last 2 weeks)
for a given region (e.g. Boston).

The infected carriers' data in the server's possession is anonymized and has already undergone redaction and transformation to remove sensitive information to limit the risk for re-identification, and  contains no personally-identifying information (PII).

Users' apps can check whether they came in contact with carrier users, and how often, while preserving privacy.  This is done with a cryptographically secure ``private set intersection'' (PSI) protocol to find matches between encrypted `point intervals' for carriers and other users. 

\item	[Step~4:] {\bf Risk assessment and notification}.
The app assesses risk for its user based on the points of contact it found via (3) and can notify users of risk. Users whose apps find them at risk due to contact with carriers can then be encouraged to get tested or self-quarantine.  The app can optionally show the user where and when the points of contact occurred.

\end{description}

This process is also illustrated above in Figure \ref{fig:high_level_schematic}. We expand upon the steps 2 and 3 in the following section.

\subsubsection{Storing and sharing GPS histories
}
As GPS points are collected by a user's device, sensitive areas are removed through either automatic redaction or manually by the user. Redaction is an important privacy step, 
%as only a handful of spatiotemporal points are enough to re-identify pseudo-anonymized users. Especially sensitive areas such as home areas can be easily inferred from where users spend time at night and subsequently cross-correlated with an external dataset like voter registrations. W
as knowledge about where someone was when, or where they commonly spend time, such as their home area, can be used to re-identify pseudo-anonymized users~\cite{golle2009anonymity,krumm2007inference}.  

The system provides two methods of redaction: automatic and manual.  
Home areas can be easily inferred by the app based on where users spend time in the nighttime, and these can be automatically redacted.  
In addition, the app can provide the user with an interface to mark additional sensitive areas for redaction.  
Any GPS point collected within an area marked for redaction is deleted and not shared.  
To further protect the user's privacy, this redaction happens on the user's device and the remaining points are then transformed or obfuscated before they are stored.
Redacting and modifying GPS points on the device, rather than after points are shared, 
is an important privacy measure to prevent users from being coerced 
into providing information on where they have been.

If a mobile app user is diagnosed as a carrier (e.g. by professional medical personnel), 
the proposed system provides two different ways for that user to 
anonymously share their GPS points to provide important information to healthcare professionals, 
other system users, and the general public.

The two GPS transformations that support these different use cases are:
\begin{enumerate}[label=(\alph*)]
    \item \label{item:analysis} GPS points are replaced by larger geographic areas that contain them to represent the areas where carriers spent time and when, without representing precise locations.
    \item \label{item:individual} Precise timestamped GPS points are transformed into ``point intervals'' and obfuscated using a one-way hash function (e.g. NIST standard SHA256 hash algorithm).
\end{enumerate}

The first way (a) is used for the aggregated view of data that we previously described as a motivating first step. The granularity level can be dialed-up or dialed-down depending on the circumstances. The more fine-grain granularity means an increase in likelihood that the user can be re-identified.

The second way (b) is used for contact tracing.  This use case is where we make a new contribution with our approach to finding points of contact while preserving privacy.

The two different use cases that \ref{item:analysis} and \ref{item:individual} serve are both central to this work as they will both be invaluable to containing the spread of infectious disease such as COVID-19.  However the remainder of this document focuses on \ref{item:individual}, as it is our main contribution and requires explanation.

\subsection{Trust and privacy model}

The proposed system is designed around a model that assumes there is a semi-trusted authority maintaining the server with diagnosed carriers' redacted location histories.  Such semi-trusted authorities could be local hospitals or local government agencies chartered and regulated to hold citizen data and maintain data privacy.
We believe that a common goal -- one that would make the proposed system usable while preserving individual privacy --
is to minimize the amount of information from a diagnosed carrier that is exposed to other users
and to the semi-trusted authority.

The proposed system is designed to minimize the amount of diagnosed carriers' information that is exposed, and to  maximize  the privacy for all other users  of the  system.  These other users  need not share any of their location  data in order  to find points of contact with  diagnosed  users. However, diagnosed users do risk forfeiting some privacy when they share their location histories with the authority managing the server \footnote{Even systems designed to be opt-in on the part of users sharing data can be abused and made compulsory by authorities once they are built.  In Singapore, people contacted by health authorities are required by law to assist in the activity mapping of their movements and interactions \cite{singapore2020trace,  tracetogetherZenDeskSayNo}.}.  Even  though they only share their redacted  and  encrypted location  data,  given  enough computational resources and malicious intent, the managing  authority can attempt to circumvent  these  measures and  reconstruct location history  data and  re-identify users\footnote{The authority managing the server with redacted and encrypted user data can attempt a grid search (brute force) attack over all possible encrypted point intervals in order to find hash collisions and reverse the one-way  hash function that encrypted the point intervals, thereby exposing the underlying data.  They can then attempt to reconstruct location histories based on the spatiotemporal correlation between data points or re-identify  users due  to the unique nature of location histories \cite{de2015unique}.}.

There is a clear need for a {\em governance model} regarding this data collection and use.  For example, data should be deleted $d$ days after it was shared, where $d$ is the number of days the diagnosed user could have transmitted the virus before sharing their location history.  There also needs to be a legal framework  in place to end the practice of collecting data in this way once the health crisis is under control\footnote{We have examples to be wary of regarding measures taken in times of crisis that extend indefinitely.  Israel declared a state of emergency during its 1948 War of Independence, justifying a range of “temporary” measures that removed individual freedoms.  They won the War of Independence but never declared their state of emergency over, and many of the “temporary” measures are ongoing \cite{harari2020the}.  Israel recently approved cellphone tracking  for its COVID-19 patients \cite{israel2020}.  Similarly COVID-19 surveillance innovations in China are likely to be used by China's counterterrorism forces beyond the pandemic to further monitor and regulate the movement of its people.  Consider the Uighur people.  The Chinese government has categorized this ethnic group as terrorists and has subjected many of them to forced labor \cite{globvoice2020}.}.

That said, we must also acknowledge that governments already have access 
to the massive amounts of location data that our proposed system would collect. 
Location histories are already collected by apps on users' smartphones, 
and the cellular towers they connect to, and through credit card purchases.

\section{Technical description}

Our proposed system broadly involves data collection followed by a method to deterministically construct hashed spatiotemporal intervals that discrete points in users' location histories are mapped to. These intervals are then used with a private set intersection protocol to inform users when points in their location histories match the points in the location histories of diagnosed carriers.
We discuss these steps in the following sections.

\begin{enumerate}
\item Collecting and Representing GPS points
\item Detecting Points of Contact Using Private Set Intersection
\item Assessing Risk and Notifying Users 
\end{enumerate}

\bigbreak

\subsection{Collecting and representing GPS points}

Timestamped GPS points are collected within user devices as they move throughout their day. These points are collected as tuples of latitude, longitude, and time:  (latitude, longitude, time). 
A user's app checks for matches between their collected points and the points shared by users who were diagnosed as carriers in order to identify points of contact. 

\subsubsection{Partition space and time into intervals}

For privacy purposes, GPS points are never directly compared in order to find these matches.
Timestamped GPS points are instead first mapped to a 3-dimensional grid, where two dimensions are for geographic space (latitude and longitude), and the third dimension is time.  We call these 3-dimensional grid cells `point intervals'.  The point intervals are then obfuscated with a deterministic one-way hash function.  Identifying points of contact becomes a matter of matching hashed point intervals.  Transforming GPS histories in this way to map (latitude, longitude, time) points that occur within a continuous spatiotemporal space into discrete `point intervals' makes comparing obfuscated GPS histories possible.  It also makes sense for our use case of finding (short) time intervals where users occupied the same spatial area\footnote{Phones collect GPS data with limited accuracy and therefore trying to match users across spatial areas with radii too small will miss points of contact.}. 

There are established ways to partition a geographic space, such as with geohashes\footnote{\url{https://en.wikipedia.org/wiki/Geohash}} for square grid cells or with the hexagonal global geospatial indexing system of H3 grid\footnote{\url{https://h3geo.org/}}. 

\begin{figure}
    \centering
    \includegraphics[width=0.8\textwidth]{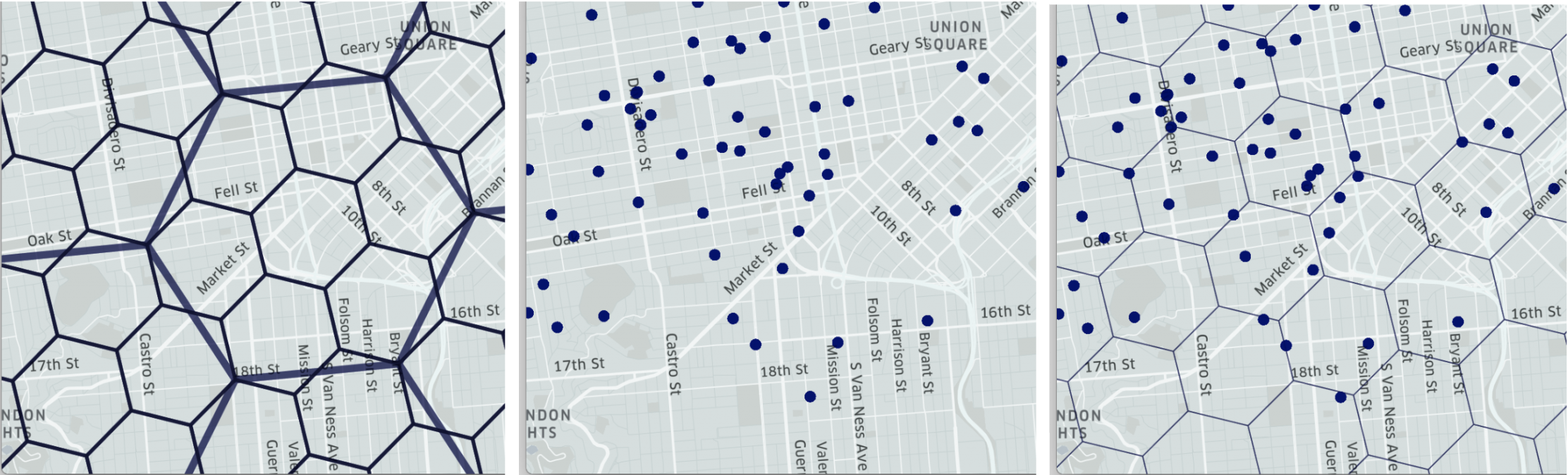}
    \caption{A geographic area partitioned by a hexagonal H3 grid.  Points are mapped to an index corresponding to their containing grid cell.}
    \label{fig:high_level_schematic2}
\end{figure}

Similarly, time can be partitioned into intervals.  For example, if an interval size is 2 minutes, then an interval boundary can always fall on the hour, and on the 2nd minute of the hour, and so on.
The 3-dimensional grid of point intervals  is an underlying system parameter (or logical ``map'') that is agreed-upon and shared across all user devices in the system.  The specific partition scheme and interval sizes used are implementation details.  What matters more is that the chosen partition scheme and the geographic and temporal resolution used are consistent across devices.

We note that when collecting GPS points there is a trade-off between accuracy in detecting points of contact and the amount of data that must be then stored and processed.  For example, if data is collected more frequently, the system is more likely to detect when users spend little time near each other, such as sharing a bus ride. However, this requires collecting and storing more data, and hence more compute resources.
We also note that the geographic partition of space can be expanded to include specific locations.  For example, a bus line might install Bluetooth beacons on its buses with unique identifiers to serve as the geographic portion of point intervals, allowing users with an app that supports this functionality to later detect if they shared a bus ride with a diagnosed carrier.

\subsubsection{Checking for matches across obfuscated GPS histories}

Since the point intervals are transformed with a one-way hash function, they cannot be easily reversed to expose underlying location histories of users.

Yet, since this transformation is deterministic, users' apps can still check for points of contact with diagnosed users by checking for matches between their transformed point intervals and those returned by a server.
Verifying whether two individuals came into contact becomes a matter of comparing 
whether the transformed point intervals  that were constructed from timestamped GPS points (latitude, longitude, time) 
collected by their devices coincide with any of the transformed point intervals
provided by the server (as well as adjacent point intervals).\footnote{Technical note about matching against adjacent intervals:
Since the partition of geographic space into intervals was predetermined, two points may be close together in space but fall into different intervals.  For this reason a user's app checks each of their collected point intervals, as well as the spatially adjacent intervals against the diagnosed carriers' shared point intervals. That is, if we use a spatial grid of hexagons (like H3), a user's collected point falls within a grid cell and that grid cell is used as an interval.  We must check that interval as well as the surrounding 6 hexagons in the grid against each data point shared by diagnosed carriers. This adds some complexity in terms of  processing power to make the comparisons $7N_u \times N_I$ rather than just $N_u \times N_I$.}

\subsection{Detecting points of contact using private set intersection}

We now describe the protocol that facilitates the private detection of matches across users' GPS histories.  
We assume that there is a semi-trusted authority  (e.g. a local health agency) operating a server.  When a patient is diagnosed as a disease carrier, they share their redacted, anonymized, hashed point intervals to the central server.  Other users' apps periodically exchange information about their own hashed point intervals with the server to detect if their hashed point intervals match against those shared by diagnosed carriers.  They do so with a private set intersection protocol. 

Private set intersection (PSI) enables two parties to compute the intersection of their data in a privacy-preserving way, such that only the common data values are revealed. It has applications in a variety of privacy sensitive settings, from measuring conversion rates for online advertising \cite{pinkas2015phasing} to securely testing sequenced human genomes \cite{baldi2011countering}.

In our case, the two parties involved are the server storing the hashed point intervals shared by diagnosed carriers, and another users' device - the client. Their data are their respective hashed point intervals. We can leverage PSI in a way so that only the user learns about the intersection of their data - the server does not learn whether it shares any point intervals in common with the user, while the users' app does learn this. Therefore, our use of PSI is designed to maximize the privacy for users who may be wary of surveillance or who do not fully trust the entities that maintain the server.

There are many PSI schemes that would fit our needs.  
These different schemes vary in their computational complexity, speed, and accuracy.
Researchers have developed fast PSI protocols optimized for a client-server model, 
including those where the client is a smartphone app~\cite{kales2019mobile} 
and where the server's dataset is significantly larger than the client's~\cite{kiss2017private}, 
which is the case for our system\footnote{  
However many of these PSI protocols achieve their improved efficiency at the cost of accuracy, 
allowing a small number of false positives, 
such as by employing bloom filters.  
This may not be an acceptable trade-off for a disease contact tracing system 
where false positives can lead to panic or the wrong people seeking scarce medical resources.}.
A good overview and comparison of PSI protocols can be found in~\cite{pinkas2015phasing}. 

To aid the reader in understanding how PSI supports our privacy goals, we provide a simple scheme using the Diffie-Hellman protocol~\cite{diffie1976new,huberman1999enhancing}
in Appendix~A.

\subsubsection{PSI benefits and implementation notes}
The use of a PSI protocol adds an extra layer of privacy protection for both the diagnosed carriers and the other users, beyond just the redaction and obfuscation of data\footnote{When other users send their point intervals to the server, their point intervals are effectively encrypted twice.  First with the deterministic hash function ($H$) that encrypts all point intervals in the same way. Second by the PSI protocol.}.

Other users only ever learn points shared by diagnosed carriers that their own points matched with (the intersection $P_I \cap P_U$).  This further protects the privacy of diagnosed carriers.  Moreover, the server need not learn whether any points match; only the other user learns the intersection of $P_I$ and $P_U$. This further protects the privacy of undiagnosed users who may be wary of their location histories leaving their device, or being shared with an authority. 

There are implementation details that further enhance privacy and efficiency. For example, servers can hold data for their local geographic regions, so that users with location histories specific to an area (e.g. the Boston area) need not interact with servers holding data specific to a far away region (e.g. the Bay Area in California). This helps subset the data so as to run PSI on a much smaller dataset, thereby helping computational efficiency. In addition, data shared by diagnosed carriers to servers should be deleted after $d$ days, as both a privacy and efficiency measure. 

The server should also limit the amount of data that a user's client device can exchange with it. It is only relevant to compare recent location histories (i.e. from the past $d$ days).  Since points are partitioned into consistent time intervals, there is therefore an upper bound on the number of points, N, that any app needs to check against the server's set of points, $P_I$. The server can limit the exchanges with any client to N points per exchange, and limit the number of queries per day.  This limitation is important for preventing privacy attacks where an adversary might attempt querying over the entire spatiotemporal grid to reconstruct the location histories of diagnosed carriers. It also reduces the computational burden for servers.

\subsection{Assessing risk and notifying users}
A user's app can assess their risk of infection based 
on the comparison (performed on the user's device)
between the point intervals on the user's device
with those received from the server. Users whose apps find them at reasonable risk can then be encouraged to get tested or self quarantine.

The implementation of our system can differ to either allow the app to learn just the number of points of contact that occured, or where and when points of contact occurred.  These different implementations have different implications for the privacy and utility that our system can offer to its users.

The number of detected points of contact is related  to how likely a user was to have spent sustained time in spaces shared with diagnosed  users,  so the number of detected points of contact is commensurate with risk and can be used to provide a risk assessment.
When the locations of points of contact are  known, the risk assessment can leverage context about these locations,  such as  whether they are confined spaces with few people versus multi-story office buildings versus outdoor parks.
Future work can further incorporate intelligence into the risk assessment.

\section{Intermediary implementation}
Given the urgency of the COVID-19 pandemic, we note that intermediary steps can be taken to implement the system we describe.  Even before a secure server is set up to perform the private set intersection (PSI) protocol, hashed point intervals for diagnosed carriers can be published to a flat data file for other users to download. While this would speed up implementation, privacy guarantees would be diminished in this case, as attackers can attempt to create points representing every potential point interval to check for matches.  Attackers could then attempt to reconstruct location histories of users diagnosed as carriers and possibly re-identify them from their shared anonymized data.  While the redaction step would decrease the likelihood of an attacker's success, some privacy risk remains. 
% Any implemented system should be opt-in so that only diagnosed users who are aware of this risk, and capable of weighing the trade-off between their individual privacy risk and the ongoing risk the pandemic poses to other users will share their data.
Finally, this intermediary implementation with a flat data file can then be subsequently transitioned to the more secure implementation using a PSI protocol.

\section{Discussion}

We proposed a technical design to address the problem of assessing users' risk of disease exposure with location histories. Our proposal is in response to existing digital contact tracing technologies, with an approach that better preserves the privacy of individuals. 

We are encouraged by other recent privacy-sensitive proposals for contact tracing \cite{canetti2020anonymous,cho2020contact,lewis2020coepi,covidwatch2020}. Some of these even extend our notions of privacy by removing the need for trust in authorities who might abuse their access to diagnosed patients' encrypted data and violate their privacy.  However, these systems are more complex and require more infrastructure and coordination, making them more difficult to implement. Our goal is to propose a system that is more privacy-preserving than the contact tracing technologies that we see governments around the world adopting, but that can also be practically implemented with the immediacy needed to both stem the spread of disease and stem the adoption of {\em privacy-violating} technologies.

Importantly, any implemented system, based on Bluetooth or GPS, should be opt-in, and clearly communicate to users how it  collects, retains, and uses data, to provide users the opportunity to weigh the trade-off between their individual privacy risk posed by sharing information with the system and the ongoing risk the pandemic poses.

In addition to the trade-offs between privacy, effectiveness, and the speed of implementation for these technologies, we must consider adoption. None of these systems can have widespread impact without extensive adoption. To expand system adoption, the proposed app's technology can be developed as an SDK (software development kit) and integrated into partnering applications that already collect user location histories, such as Google Maps. These partner applications can then ask the user for the extra permissions and consent for this system's use case.  There are many such applications that already collect user location histories in the background.  They often use this information to serve the user more relevant advertisements and content.  This data is used to improve the user  experience but more often serves private profit.  Now, in the face of the COVID-19 pandemic, is the time for industry and researchers to come together and for the ubiquitous collection of location data to serve the public good.
At the same time, we must avoid building any new surveillance systems to serve as short-term emergency measures that can last beyond a time of crisis.

\section{Acknowledgements} 
We thank Thomas Hardjono, Dan Calacci, Ethan Zuckerman, and Robert Obryk for contributing their ideas and critiques, and we thank Connection Science and Media Lab sponsors for their support. 
\bibliographystyle{unsrt}
\bibliography{covid}

\begin{thebibliography}{10}

\bibitem{lewis2020coepi}
Dana~M. Lewis et~al.
\newblock Coepi: Community epidemiology in action.
\newblock \url{https://github.com/Co-Epi}, note = {Accessed: 2020-03-30}, 2020.

\bibitem{singapore2020trace}
Government~Digital Services and Blue Trace.
\newblock Tracetogether.
\newblock \url{https://www.tracetogether.gov.sg/}, note = {Accessed:
  2020-03-30}, 2020.

\bibitem{covidwatch2020}
Rhys Fenwick, Mike Hittle, Mark Ingle, Oliver Nash, Victoria Nguyen, James
  Petrie, Jeff Schwaber, Zsombor Szabo, Akhil Veeraghanta, Mikhail Voloshin,
  Sydney Von~Arx, and Tina White.
\newblock Covid watch.
\newblock \url{https://www.covid-watch.org/}, note = {Accessed: 2020-03-30},
  2020.

\bibitem{raskar2020apps}
Ramesh Raskar, Isabel Schunemann, Rachel Barbar, Kristen Vilcans, Jim Gray,
  Praneeth Vepakomma, Suraj Kapa, Andrea Nuzzo, Rajiv Gupta, Alex Berke, et~al.
\newblock Apps gone rogue: Maintaining personal privacy in an epidemic.
\newblock {\em arXiv preprint arXiv:2003.08567}, 2020.

\bibitem{singaporeCases}
Government~Digital Services.
\newblock Covid-19: Cases in singapore.
\newblock \url{https://www.gov.sg/article/covid-19-cases-in-singapore}, note =
  {Accessed: 2020-03-30}, 2020.

\bibitem{zastrow2020south}
M~Zastrow.
\newblock South korea is reporting intimate details of covid-19 cases: has it
  helped?
\newblock {\em Nature}, 2020.

\bibitem{nytimesCoronaVirusCount}
The New~York Times.
\newblock Coronavirus in the u.s.: Latest map and case count.
\newblock
  \url{https://www.nytimes.com/interactive/2020/us/coronavirus-us-cases.html},
  note = {Accessed: 2020-03-30}, 2020.

\bibitem{johnhopkinsCOVID19cases}
John~Hopkins University and Medicine.
\newblock Coronavirus covid-19 global cases.
\newblock \url{https://coronavirus.jhu.edu/map.html}, note = {Accessed:
  2020-03-30}, 2020.

\bibitem{nebraskaCOVID19}
Nebraska~Department of~Health and Human Services.
\newblock Coronavirus disease 2019 (covid-19).
\newblock \url{http://dhhs.ne.gov/Pages/Coronavirus.aspx}, note = {Accessed:
  2020-03-30}, 2020.

\bibitem{haaretzIsraelApprovesTracking}
Haaretz.
\newblock Israel approves cellphone tracking for coronavirus patients as cases
  rise to 213.
\newblock
  \url{https://www.haaretz.com/israel-news/.premium-israel-approves-cellphone-tracking-for-coronavirus-patients-as-cases-rise-to-213-1.8678361},
  note = {Accessed: 2020-03-30}, 2020.

\bibitem{nytimesChinaColorCode}
The New~York Times.
\newblock In coronavirus fight, china gives citizens a color code, with red
  flags.
\newblock
  \url{https://www.nytimes.com/2020/03/01/business/china-coronavirus-surveillance.html},
  note = {Accessed: 2020-03-30}, 2020.

\bibitem{de2015unique}
Yves-Alexandre De~Montjoye, Laura Radaelli, Vivek~Kumar Singh, et~al.
\newblock Unique in the shopping mall: On the reidentifiability of credit card
  metadata.
\newblock {\em Science}, 347(6221):536--539, 2015.

\bibitem{niyogi2020memo}
Sourabh Niyogi, James Petrie, Scott Leibrand, Jack Gallagher, Hamish, Manu
  Eder, Zsombor Szabo, and George Danezis.
\newblock Cen proposals for privacy-preserving distributed contact tracing.
\newblock
  \url{https://docs.google.com/document/d/1f65V3PI214-uYfZLUZtm55kdVwoazIMqGJrxcYNI4eg},
  note = {Accessed: 2020-03-30}, 2020.

\bibitem{niyogi2020github}
Sourabh Niyogi, Scott Leibrand, and Dana~M. Lewis.
\newblock Cen proposals for privacy-preserving distributed contact tracing.
\newblock \url{https://github.com/Co-Epi/coepi-backend-go}, note = {Accessed:
  2020-03-30}, 2020.

\bibitem{cho2020contact}
Hyunghoon Cho, Daphne Ippolito, and Yun~William Yu.
\newblock Contact tracing mobile apps for covid-19: Privacy considerations and
  related trade-offs.
\newblock {\em arXiv preprint arXiv:2003.11511}, 2020.

\bibitem{kampf2020persistence}
G{\"u}nter Kampf, Daniel Todt, Stephanie Pfaender, and Eike Steinmann.
\newblock Persistence of coronaviruses on inanimate surfaces and its
  inactivation with biocidal agents.
\newblock {\em Journal of Hospital Infection}, 2020.

\bibitem{golle2009anonymity}
Philippe Golle and Kurt Partridge.
\newblock On the anonymity of home/work location pairs.
\newblock In {\em International Conference on Pervasive Computing}, pages
  390--397. Springer, 2009.

\bibitem{krumm2007inference}
John Krumm.
\newblock Inference attacks on location tracks.
\newblock In {\em International Conference on Pervasive Computing}, pages
  127--143. Springer, 2007.

\bibitem{tracetogetherZenDeskSayNo}
Team~Trace Together.
\newblock Can i say no to uploading my tracetogether data when contacted by the
  ministry of health?
\newblock
  \url{https://tracetogether.zendesk.com/hc/en-sg/articles/360044860414-Can-I-say-no-to-uploading-my-TraceTogether-data-when-contacted-by-the-Ministry-of-Health-},
  note = {Accessed: 2020-03-30}, 2020.

\bibitem{harari2020the}
Yuval~Noah Harari.
\newblock The world after coronavirus.
\newblock {\em Financial Times}, March 2020.

\bibitem{israel2020}
Oliver Holmes.
\newblock Israel to track mobile phones of suspected coronavirus cases.
\newblock {\em The Guardian}, March 2020.

\bibitem{globvoice2020}
Shui-yin~Sharon Yam.
\newblock Coronavirus and surveillance tech: how far will gov’ts go and will
  they stay when they get there?
\newblock {\em Hong Kong Free Press}, March 2020.

\bibitem{pinkas2015phasing}
Benny Pinkas, Thomas Schneider, Gil Segev, and Michael Zohner.
\newblock Phasing: Private set intersection using permutation-based hashing.
\newblock In {\em 24th $\{$USENIX$\}$ Security Symposium ($\{$USENIX$\}$
  Security 15)}, pages 515--530, 2015.

\bibitem{baldi2011countering}
Pierre Baldi, Roberta Baronio, Emiliano De~Cristofaro, Paolo Gasti, and Gene
  Tsudik.
\newblock Countering gattaca: Efficient and secure testing of fully-sequenced
  human genomes (full version).
\newblock {\em arXiv preprint arXiv:1110.2478}, 2011.

\bibitem{kales2019mobile}
Daniel Kales, Christian Rechberger, Thomas Schneider, Matthias Senker, and
  Christian Weinert.
\newblock Mobile private contact discovery at scale.
\newblock In {\em 28th $\{$USENIX$\}$ Security Symposium ($\{$USENIX$\}$
  Security 19)}, pages 1447--1464, 2019.

\bibitem{kiss2017private}
{\'A}gnes Kiss, Jian Liu, Thomas Schneider, N~Asokan, and Benny Pinkas.
\newblock Private set intersection for unequal set sizes with mobile
  applications.
\newblock {\em Proceedings on Privacy Enhancing Technologies},
  2017(4):177--197, 2017.

\bibitem{diffie1976new}
Whitfield Diffie and Martin Hellman.
\newblock New directions in cryptography.
\newblock {\em IEEE transactions on Information Theory}, 22(6):644--654, 1976.

\bibitem{huberman1999enhancing}
Bernardo~A Huberman, Matt Franklin, and Tad Hogg.
\newblock Enhancing privacy and trust in electronic communities.
\newblock In {\em Proceedings of the 1st ACM conference on Electronic
  commerce}, pages 78--86, 1999.

\bibitem{canetti2020anonymous}
Ran Canetti, Ari Trachtenberg, and Mayank Varia.
\newblock Anonymous collocation discovery:taming the coronavirus while
  preserving privacy, 2020.

\bibitem{chen2017fast}
Hao Chen, Kim Laine, and Peter Rindal.
\newblock Fast private set intersection from homomorphic encryption.
\newblock In {\em Proceedings of the 2017 ACM SIGSAC Conference on Computer and
  Communications Security}, pages 1243--1255, 2017.

\end{thebibliography}

\renewcommand\thefigure{\thesection.\arabic{figure}}    
\setcounter{figure}{0}  

\newpage
\appendix
\section{An example private set intersection protocol using Diffie-Hellman}
Below we outline an example of how our system would work with a simple Diffie-Hellman PSI protocol. Note that the actual implementation may differ from this example. Our description assumes familiarity with Diffie-Hellman, modular arithmetic and concepts from cryptography such as the discrete log problem. Readers can otherwise skip this section to the PSI protocol summary below. 

Before we walk through this PSI protocol, we clarify the problem and notation.

\paragraph{Notation and Problem Statement:}

We call a point interval $p$, and a collected sequence of point intervals $P = [p_1, p_2, ...]$.  
We call the users' point intervals that are collected by their device $P_U$. 
We call the point intervals collected for diagnosed carriers and later shared with the server $P_I$.

As we noted earlier, each point interval is encrypted by a commonly shared deterministic hash function, which we call $H$.  This means that a user's  phone really stores $H(P_U) = [H(p_{U1}), H(p_{U1}), ... H(p_{Un})]$, and the server stores \\${H(P_I) = [H(p_{I1}), H(p_{I1}), ... H(p_{Im})]}$.

If a user has a point interval matching one shared by a diagnosed carrier, i.e. $p_{Ui} = p_{Ij}$ for some $p_{Ui}$ in $P_U$ collected by the user's device, and some $p_{Ij}$ in $P_I$ collected by a diagnosed carriers' device and later shared to the server, then the encrypted hashes of these point intervals match as well.  $H(p_{Ui}) = H(p_{Ij})$.

Consider $H(P_U)$ as a set which is stored on the user's device, and $H(P_I)$ as a set stored on the server.  The problem is then to allow the user's app to learn the set intersection of $H(P_U)$ and $H(P_I)$.

\paragraph{Protocol:}
Our described use of Diffie-Hellman is written with the multiplicative group of integers modulo $p$, where $p$ is prime, and $g$ is a primitive root modulo $p$.

\paragraph{Setup:} The server and the client user's device have an agreed upon modulus, $p$, and base of the  multiplicative group, $g$.
The server and client each generate secret private keys, $a$ and $b$, respectively.

\begin{enumerate}
    \item The client encrypts the user's hashed point intervals, $H(P_U)$, with a and sends this data to the server.
    
    $Client \rightarrow Server : H(P_U)^a = [H(p_{U1})^a, H(p_{U1})^a, ..., H(p_{Un})^a] \bmod p$
    \item The server encrypts its stored hashed point intervals, $H(P_I)$, with $b$ and sends this data to the client.
    
    $Server \rightarrow Client : H(P_I)^b = [H(p_{I1})^b, H(p_{I1})^b, ... H(p_{Im})^b] mod  p$
    
    \item Upon receiving the encrypted hashed point intervals sent by the client $H(P_U)^a$, the server further encrypts this data with its key b and sends the result  back to the client.
    
    $Server \rightarrow Client: H(P_U)^{ab} = [H(p_{U1})^{ab}, H(p_{U1})^{ab}, ..., H(p_{Un})^{ab}] \bmod p$
    
    \item The client receives both $H(P_I)^b$ and $H(P_U)^{ab}$.  The client then further encrypts $H(P_I)^b$ with its key a to create 
    
    $H(P_I)^{ba} = [H(p_{I1})^{ba}, H(p_{I1})^{ba}, ..., H(p_{Im})^{ba}] \bmod p$.
    
    \item The client can then compute the set intersection by comparing the elements of $H(P_U)^{ab}$ and $H(P_I)^{ba}$.

\end{enumerate}

Due to the multiplicative properties of the group, any matching $H(p_U)$ and $H(_I)$ values will have matching $H(p_U)^{ab}$ and $H(p_I)^{ba}$ values. This means that if the client has any point intervals that match point intervals shared to the server, $p_U = p_I$, then $H(p_U)^{ab} =  H(p_I)^{ba}$, and these matches will be detected by the client.

\begin{figure}
    \centering
    \includegraphics[width=0.8\textwidth]{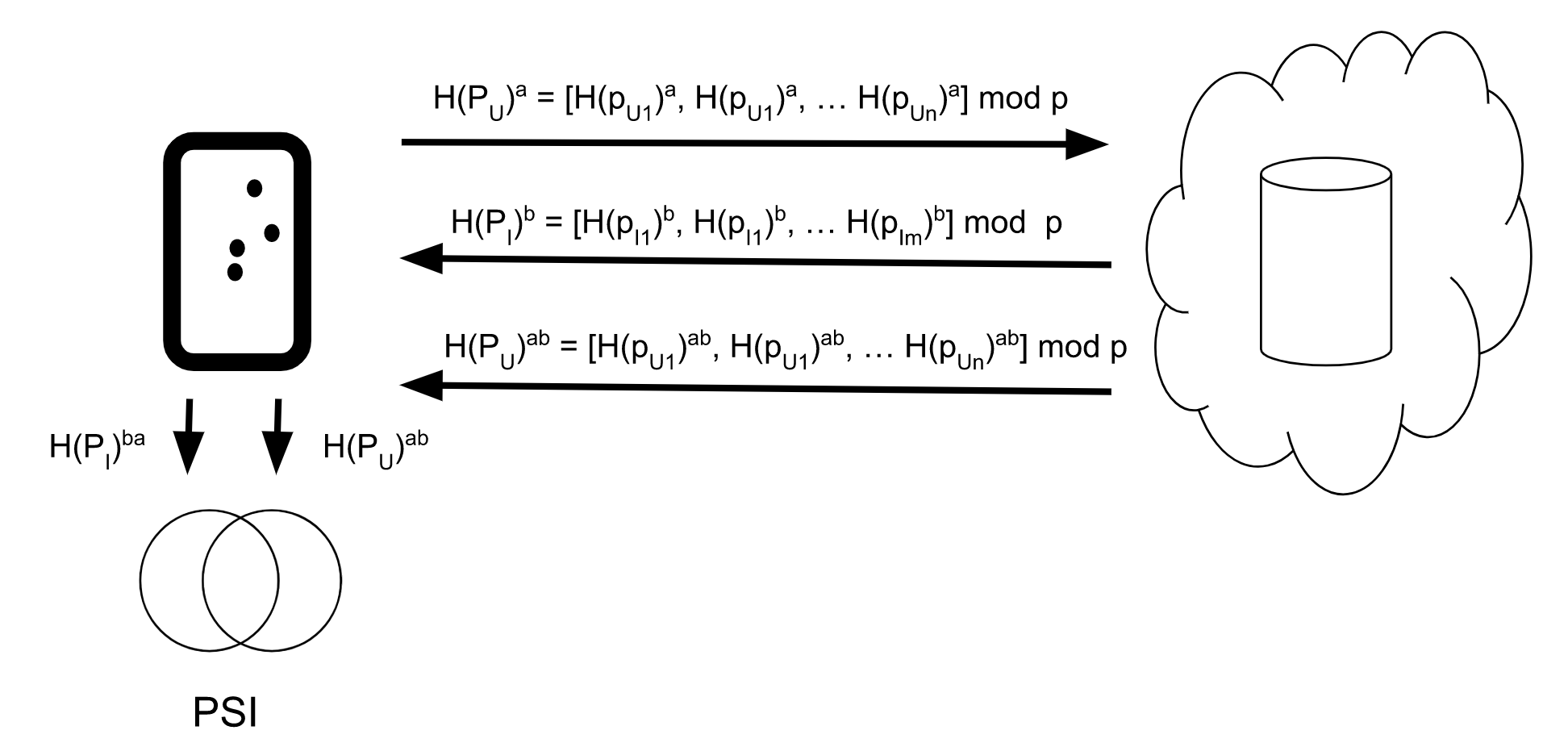}
    \caption{This figure shows our private set intersection (PSI) protocol based on Diffie-Hellman key exchange. The intersection is done on encrypted point intervals corresponding to the client's and carrier's location traces. The encrypted point intervals from diagnosed carriers are stored in a server. Only the client device learns the intersection of data, where the intersection is the set of point intervals it has in common with the point intervals on the server $(P_U \cap P_I)$.  Points in $(P_U \cap P_I)$ are represented in $(H(P_U)^{ab}\cap H(P_I)^{ba})$.}
    \label{fig:high_level_schematic3}
\end{figure}

This protocol allows for flexibility in terms of whether it allows a client to learn which of its points have matches versus how many of its points have matches.  At step (3) of the protocol the server further encrypts the data received from the client, H(PU)a, and returns H(PU)ab.  If the server maintains the order in the sequence of points intervals, then the client can then learn exactly which hashed point intervals in the sequence it sent to the server, $H(P_U)^a = [H(p_{U1})^a, H(p_{U1})^a, ... H(p_{Un})^a]$ match against items in the server's encrypted data, $H(P_I)^{ba}$.  If the server instead shuffles the sequence before returning it in step (3), then the client can learn how many of its point intervals match against the server's data, but not which ones do.  

We described a simple protocol in order to more easily explain how our system can operate.  Optimizations can be made for efficiency.  For example, the server can reuse its private key, b, and set of encrypted data across multiple interactions with different clients.  It can refresh this key and re-encrypt its data periodically, or as new data is shared by diagnosed carriers or deleted as it becomes old.  Decreasing how often the server encrypts its data can increase its efficiency.
Faster PSI protocols have been developed, including those that optimize for the exchange of information between a server and client, particularly where the server has a much larger set of data than the client, such as in our use case \cite{chen2017fast,kiss2017private}.

\end{document}